\documentclass[journal,onecolumn]{IEEEtran}
\IEEEoverridecommandlockouts
\usepackage{cite}
\usepackage{amsthm,amsmath,amssymb,amsfonts}
\usepackage{algorithmic}
\usepackage{graphicx}   
\usepackage{textcomp}
\usepackage{xcolor}
\usepackage{lipsum}
\usepackage[utf8]{inputenc}
\def\BibTeX{{\rm B\kern-.05em{\sc i\kern-.025em b}\kern-.08em
    T\kern-.1667em\lower.7ex\hbox{E}\kern-.125emX}}
\usepackage{hyperref}

\theoremstyle{definition}
\newtheorem{example}{Example}[section]

\begin{document}

\title{Teaching Programming in the Age of Generative AI: Insights from Literature, Pedagogical Proposals, and Student Perspectives}

\author{
\IEEEauthorblockN{Clemente Rubio-Manzano (1), Jazna Meza (1,2),  Rodolfo Fernández-Santib\'añez (1,2), Christian Vidal-Castro (1)}
\IEEEauthorblockA{(1)\textit{Departamento de Sistemas de Información} -
\textit{Universidad del B\'io-B\'io} -
Concepci\'on, Chile}
\IEEEauthorblockA{(2)\textit{ Escuela de Informática y Telecomunicaciones} -
\textit{DuocUC} -
Concepci\'on, Chile}
}

\maketitle

\begin{abstract}
Computer programming is undergoing a true transformation driven by powerful new tools for automatic source code generation based on large language models. This transformation is also manifesting in introductory programming courses at universities around the world, generating an in-depth debate about how programming content should be taught, learned, and assessed in the context of generative artificial intelligence.

This article aims, on the one hand, to review the most relevant studies on this issue, highlighting the advantages and disadvantages identified in the specialized literature. On the other hand, it proposes enriching teaching and learning methodologies by focusing on code comprehension and execution rather than on mere coding or program functionality. In particular, it advocates for the use of visual representations of code and visual simulations of its execution as effective tools for teaching, learning, and assessing programming, thus fostering a deeper understanding among students.

Finally, the opinions of students who took the object-oriented programming course are presented to provide preliminary context supporting the incorporation of visual simulations in Java (or other languages) as part of the training process.

\end{abstract}


\section{Introduction} 

Computer programming is undergoing a true revolution, and its teaching is experiencing an accelerated transformation due to the rise of new generative artificial intelligence tools ($\text{IAGen}$\footnote{The acronym $\text{IAGen}$ is used to distinguish it from Artificial General Intelligence, not $\text{IAG}$.}). These tools, based on large language models ($\text{LLMs}$), have proven to be highly effective in addressing a wide range of programming-related tasks. They are not only capable of explaining error messages but also of proposing viable solutions to them. This capability has generated intense debate about the future of programming and the role of human programmers, giving rise to new trends such as "vibe coding," which promotes the development of applications without requiring prior programming knowledge or understanding of the source code \cite{maes2025gotchas}.
In this context, one of the main concerns associated with the use of $\text{IAGen}$ is plagiarism. This concern has been widely highlighted by teachers who assess their students through written assignments. The ability of $\text{LLMs}$ to generate high-quality texts makes it considerably difficult to determine whether the content presented was actually created by the students themselves, despite the emergence of recent tools for detecting texts generated by artificial intelligence. A similar problem is observed in introductory programming courses, as well as in more advanced ones, where the concern focuses on the improper use of automatic code generators. In these environments, it is increasingly difficult to distinguish between genuine learning and the mere delegation of tasks to systems based on generative AI.

To address this issue, research has recently begun on the impact of using $LLMs$ on programming education. For example, \cite{boguslawski2025programming} presents one of the first studies exploring the relationship between student motivation and the use of $LLMs$ in introductory programming courses. The study reveals that students already make extensive use of these tools, especially for bug fixing, code debugging, and gaining a deeper understanding of fundamental programming concepts. Meanwhile, \cite{becker2023generative} argues that interactive tools based $IAGen$ have enormous transformative potential for initial programming education, with direct implications for the curriculum of computer science programs, affecting key issues such as what is taught, when, how, and to whom. However, this remains a contested area: while some teachers are beginning to actively incorporate $IAGen$ into their teaching practices, others express reservations, fearing that the challenges outweigh the benefits. In this context, it is urgent to design teaching strategies that critically integrate these tools, balancing their advantages with active learning approaches that foster critical thinking and deep understanding.
The objective of this paper is to conduct a preliminary review of the main studies addressing this issue, with special emphasis on the advantages and disadvantages identified in the literature. Based on this analysis, we propose a shift in the approach to teaching and assessing introductory programming courses. In particular, we argue that traditional methodologies should be complemented with approaches focused on code comprehension, beyond mere writing or execution. Within this framework, we advocate for the use of visual representations of code and visual simulations of program execution as key pedagogical tools to assess students' actual comprehension.
The main reason for proposing this methodological extension lies in the limited capacity for real understanding demonstrated by $LLMs$ when generating code. In simple terms, these models do not truly understand what they are programming. While they can produce syntactically correct code and even generate coherent explanations about its operation, these explanations are not based on a semantic understanding of the code but rather on statistical correlations derived from the syntactic structures present in large volumes of data. Currently, $LLMs$ lack computational thinking in the strict sense; they do not build abstract mental representations that allow them to reason about the execution of a program. In other words, an $IAGen$ does not possess cognition \cite{shojaee2025illusion,valmeekam2022large}.

The rest of the article is organized as follows: Section \ref{sec-ventajas-desventajas} presents the main advantages and disadvantages identified in the specialized literature regarding the use of large language models in teaching and learning programming. Section \ref{sec-comprension-codigo} addresses, in detail, the importance of code comprehension and visual program simulation, justifying why we believe that teaching, learning, and assessment processes should be based on visualization techniques. It also discusses how these strategies can be incorporated into introductory programming courses. Section \ref{sec-propuesta} describes the assessment proposal based on the visualization of program execution, aimed at assessing code comprehension during the training process. Finally, Section \ref{sec-final} presents the conclusions of the work and points out possible lines for future research.

\section{Advantages and Disadvantages of Using Large Language Models to Teach Programming}
\label{sec-ventajas-desventajas}

Recent literature has highlighted both the advantages and limitations of using $IAGen$ based on $LLMs$ in programming education. 

\subsection{Advantages}

\begin{enumerate}

\item The use of this technology can enhance students' autonomy and skill development by facilitating personalized learning. Additionally, it contributes to increased motivation and reduced frustration levels during the learning process. \cite{boguslawski2025programming}

\item From a teaching perspective, $IAGen$ has proven capable of generating novel problems and examples, including correct solutions and functional test cases. Furthermore, it can serve as an assessment tool, providing automatic and timely feedback. Its potential as a virtual teaching assistant can alleviate the workload of both teachers and their teaching assistants \cite{becker2023generative}.
\item 
Other studies highlight various applications of $IAGen$, such as personalized tutoring, knowledge reinforcement, the development of teaching materials, the generation of source code, the delivery of immediate feedback, and support in evaluation processes \cite{garcia2025teaching}.
\item  Preferences for $IAGen$ tools in problem-solving and conceptual understanding are also mentioned. Overall, participants benefit the most from the interpretation, analysis, and evaluation dimensions of critical thinking. \cite{clarke2025impact}
\item  The findings also suggest that teachers can develop more structured learning approaches with $IAGen$, utilizing reflection and self-regulation more explicitly \cite{clarke2025impact}.
\end{enumerate}

\subsection{Disadvantages}

\begin{enumerate}
\item Current systems are currently unable to effectively provide or replace social support, which remains a key factor in student motivation and engagement \cite{boguslawski2025programming}.
\item There is a risk that students may develop an overdependence on technology, which could negatively impact their learning autonomy. Furthermore, unequal access to these tools—stemming from differences in subscription plans—could exacerbate existing gaps between students \cite{becker2023generative}.
\item The use of generative artificial intelligence in teaching poses significant risks, including academic dishonesty and ethical dilemmas. Additionally, there is a potential decline in the development of critical thinking skills, which aligns with the overreliance on and various technical limitations of these tools \cite{garcia2025teaching}.
\item This reliance highlights the importance of appropriate guidance from teachers to foster genuine critical thinking in students \cite{clarke2025impact}.
\item Moreover, it has been observed that students may benefit less from essential skills such as inference, explanation, and self-regulation. If not carefully integrated, these technologies could compromise student autonomy and limit overall development \cite{clarke2025impact}.

\end{enumerate}

\section{Understanding code and visually stimulating programs}
\label{sec-comprension-codigo}

\begin{figure}[t]
\centering
\includegraphics[width=12.5cm]{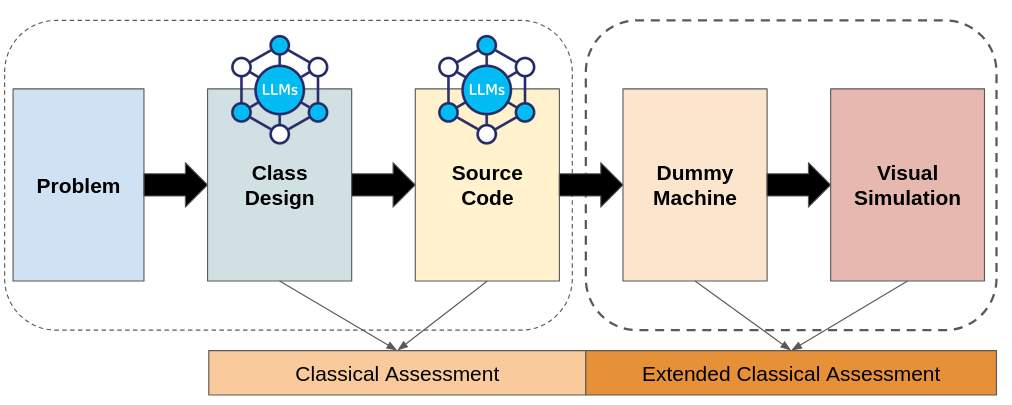}
\caption{Traditionally, assessment in programming courses has primarily focused on lesson design and implementation, emphasizing code functionality rather than student understanding. In this approach, key aspects such as the mental representation of how the program works are often neglected. Therefore, we propose incorporating two additional elements into the assessment process: the use of a dummy machine and visual simulation, to support and demonstrate students' understanding of the source code.}
\label{fig-nueva-evaluacion}
\end{figure}

In our view, the central debate should not focus on prohibiting the use of automatic code generation tools. We believe that such tools can be valuable educational resources that, when properly targeted, can be highly beneficial. However, it is essential to expand and rethink the teaching-learning and assessment processes (see Figure \ref{fig-nueva-evaluacion}). It is necessary to go beyond merely assessing the submitted code and its functionality, incorporating as a fundamental axis the students' understanding of the code they present.

In this context, visualization is presented as a particularly intuitive technique for understanding and is recognized as a highly effective educational tool \cite{escudero2012modelos}. There is also broad consensus among education research specialists regarding the recommendation to use visualizations in learning environments. The use of images to illustrate and clarify the functioning of various processes is a widespread teaching practice, present in both textbooks and classroom environments around the world, extending beyond the specific scope of computer science education \cite{de2025modelo}.

It should be noted that the educational value of visualization is not a recent idea. As early as the 1980s, Mayer advocated for its use as a conceptual model to facilitate the learning of a fictitious machine \cite{mayer1981psychology}. Later, the same author developed a theory of multimedia learning that highlights the importance of using visual channels as a complementary route to understanding \cite{mayer2005cambridge}. Along these same lines, various studies have argued that if we assume human cognition is partly visual in nature, then the use of visualizations should have a positive impact on learning processes \cite{ben2001constructivism}.

From this perspective, visualization constitutes a valuable tool for highlighting variations in critical aspects of a phenomenon to be learned. Indeed, Marton and colleagues have shown that a visualization system designed for teaching Newtonian physics can facilitate shifts in students' perspectives, promoting a deeper understanding of the content. It has also been suggested that visualization can be especially useful for emphasizing fundamental aspects of key programming concepts, such as objects and variables.

Tools such as Python Tutor \cite{guo2013online} have proven particularly successful in this area, leading to various investigations focused on the study of visual representations and their impact on learning \cite{balasubramanian2024challenges}. More recently, visualization has gained increasing importance in the teaching and learning of algorithms. In this context, proposals have been developed that incorporate interactive visualizations to enhance the understanding of fundamental algorithms, such as search and sorting \cite{singh2024sorting}.

\subsection{Incorporating code understanding and visualization techniques within programming curricula}

The Visual Program Simulation (VPS) technique seeks to actively involve students in the simulated execution of programs, assuming the role of the computer. By visualizing a fictitious machine—an abstract representation of how a computer works—students can observe and interpret what happens in memory as the code executes. The objective of these simulations is to help beginners in programming develop the ability to reason about the dynamic behavior of programs, a skill identified as particularly challenging in introductory programming courses. VPS promotes meaningful learning by encouraging students' cognitive participation through visual representations of the execution process.

\section{Our Proposal: Emphasizing the Understanding of Program Execution}
\label{sec-propuesta}

Teaching programming presents a considerable challenge during the early years of computer science studies. This difficulty largely stems from the high demand for abstraction and imagination skills necessary to understand the internal processes involved in developing and executing a program. Based on our experience accumulated over more than 25 years in the field of computer programming—15 of which have been dedicated to teaching—we believe that the prohibition on the use of generative artificial intelligence (IAGen) tools is inappropriate. However, we maintain that it is essential to rethink current approaches to teaching, learning, and assessing programming. In the current context, the focus should not be exclusively on the final product—the delivered code—but rather on the student's deep understanding of that code. In this sense, fostering students' ability to explain what they have programmed, particularly to their peers, could constitute an effective pedagogical strategy for promoting meaningful learning.

Consequently, one of the most relevant challenges in teaching programming is identifying which techniques, or combinations of them, are most effective in fostering comprehension and the ability to explain code. Our teaching experience leads us to affirm that, in the current context, it is essential for students to develop skills such as understanding the execution process of a program, identifying and debugging errors, logically following execution step by step, and constructing abstract mental representations of code behavior. These competencies emerge as key to successfully addressing new scenarios in the teaching and learning of programming.

\subsection{Visual Representation of Java Programs}

We propose using visual representations of objects during runtime as a strategy to facilitate program comprehension. These representations effectively support the analysis of source code execution. In our proposal, this technique is integrated into a problem-based learning methodology, resulting in a teaching approach that combines problem-solving with an assessment focused on code comprehension.

Process Steps is as follows: First, the problem to be solved is defined, which requires the implementation of a class with specific attributes and at least one functional method. The student can choose to develop the code independently or utilize a generative artificial intelligence tool (referred to as $IAGen$). The source code is then reviewed to detect potential syntax or logic errors. Once the code is validated, it is analyzed, beginning with a visual representation of the objects involved and continuing with a simulation of their execution, in order to foster a deeper understanding of the program's behavior.

To visually simulate program execution, it is necessary to design a fictitious machine that serves as a conceptual model of the execution environment. In our case, this machine is composed exclusively of boxes and arrows, which represent data and interactions between objects through virtual references or pointers. Depending on the type of data involved, these representations can range from simple structures to more complex configurations. Based on this visual logic, the following types can be distinguished:

\begin{itemize}
\item Primitive type.
\item Array type.
\item Object type with primitive type attributes.
\item Array of objects with primitive type attributes.
\item Array of objects with object type attributes.
\end{itemize}

\subsubsection{Simple Types (primitive, array and objects)}

\begin{figure}[t]
\centering
\includegraphics[width=15.5cm]{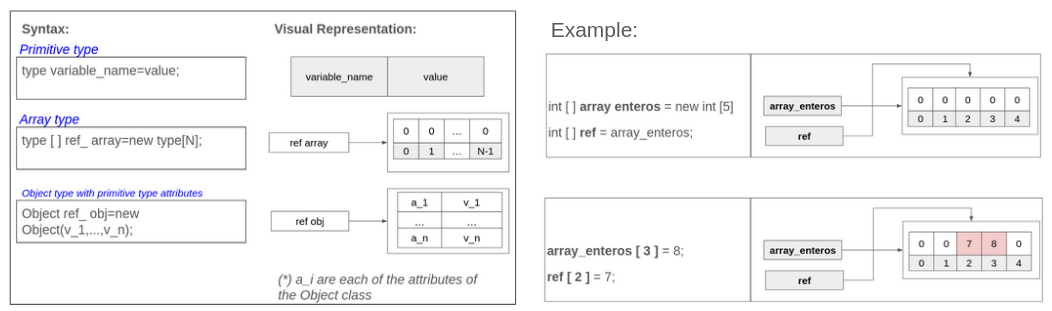}
\caption{On the right, Visual representation of primitive types, the $array$ type and the object type with primitive type attributes. On the left, an example for a visual representation of the creation of an $array$ and two references pointing to the memory area where it was saved and visual representation of access to $array$ through the two references that point to $array$.}
\label{fig-representacion-simples}
\end{figure}

For primitive types, two boxes are used: one containing the name of the variable and the other its value. For the $array$ type, one box holds the name of the $array$, another box contains the positions and values of the $array$, and an arrow indicates the relationship between the $array$ name and the memory space where the $array$ is stored (see Figure \ref{fig-representacion-simples}). In other words, there is a designated memory area for the $array$, and there may be references pointing to that area.

\begin{example}
Let's assume an $array$ of integers of size $5$. The visual representation would be as shown in Figure \ref{fig-representacion-simples}.
That is, “$array\_enteros$” is a reference that points to a memory area where the $array$ structure is located (boxes indexed by a number). As can be seen when establishing the assignment between references “$ref=array\_enteros$”, the reference “$ref$” now points to the same location as “$array\_enteros$”. Through this reference, along with the brackets and the position, you can access the elements of the $array$. For example, in Figure \ref{fig-representacion-simples}, you can see how to access the $array$ using the reference “$array\_enteros$” and the reference “$ref$”.

\end{example}

\begin{example}
On the other hand, the visual simulation of the creation of objects derived from the classes we designed and implemented is essential. Let's assume a $Person$ class composed of two public attributes ($rut$ and $age$), along with its constructor. Additionally, we have the $Main$ class, which contains the $main$ method; this method is essential for starting the execution of a Java program.

\begin{figure}[h]
\centering
\includegraphics[width=10.5cm]{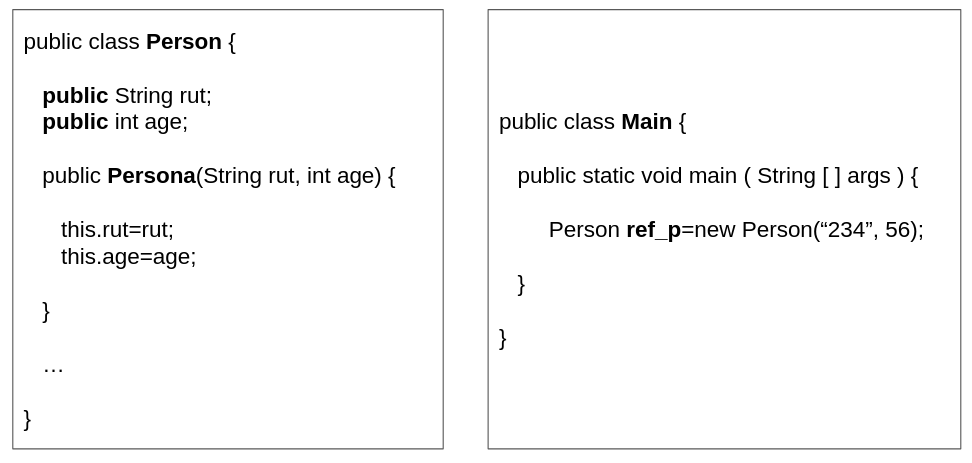}
\caption{$Person$ class with public attributes to help understand access to object attribute values.}
\label{fig-ejemplo-2a}
\end{figure}

What happens during the execution of $main$ involves two main actions (see Figure \ref{fig-ejemplo-2b}). First, a reference called "$ref\_p$'' is created that points to the memory area where an object of type "$Person$" is stored. This type of object receives two values: the "$rut$" of the person (identifier) and the "$age$". In this case, the string "$234$" is stored in the "$rut$" attribute, while the value "$56$" is stored in "$age$2. These values can be accessed through the reference "$ref\_p$". For example, you could run "$ref\_p.rut$" to obtain the string "$234$" and "$ref\_p.edad$'' to access the age of "$56$". Note that you can make one reference point to another (e.g., "$ref2 = ref\_p$"), meaning you can access the object pointed to by "$ref\_p$" through "$ref2$". Consequently, if we modify the value of "$rut$" to “$000$” through "$ref2$", it will also be visible when accessed by "$ref\_p$" (see Figure \ref{fig-ejemplo-2c}).

\begin{figure}[ht]
\centering
\includegraphics[width=10.5cm]{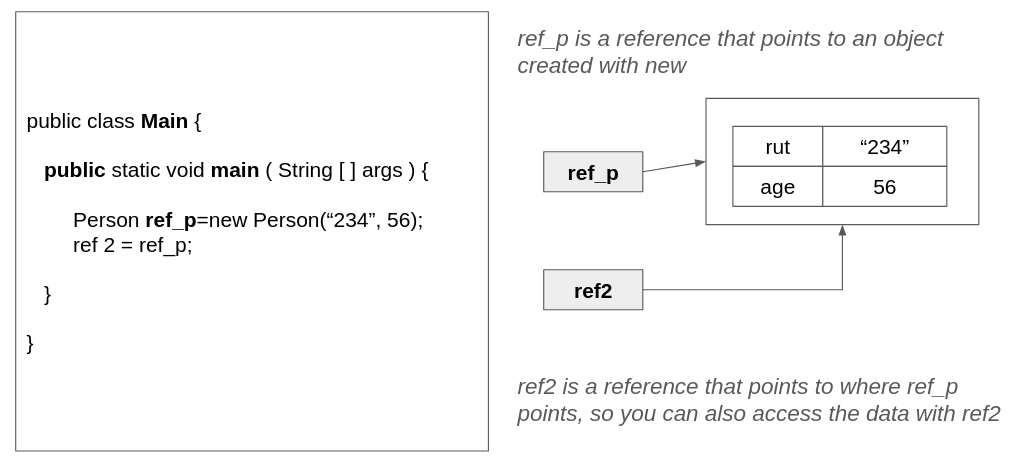}
\caption{Visual execution simulation consists of simulating step by step the execution of the instructions written in the $main$ method.}
\label{fig-ejemplo-2b}
\end{figure}

\begin{figure}[ht]
\centering
\includegraphics[width=10.5cm]{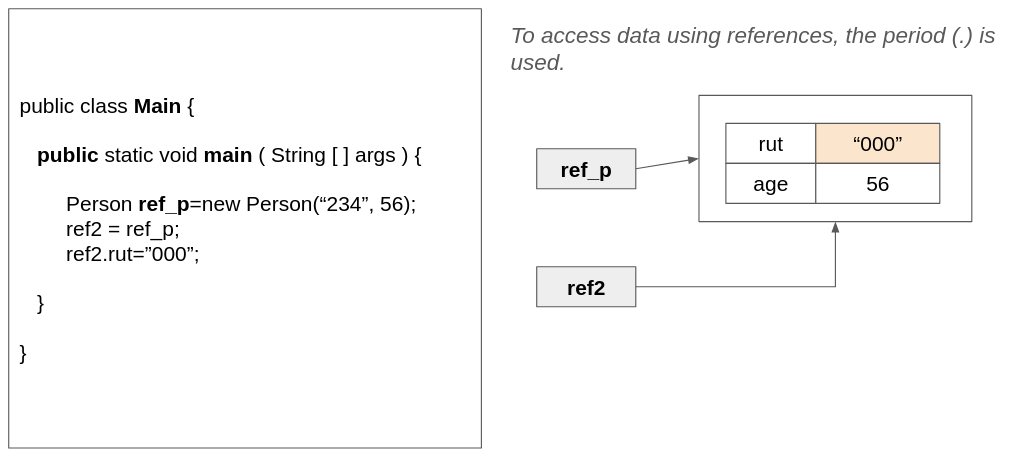}
\caption{The visual simulation of the execution of this code shows how a reference a is created that points to an object ($new$). In addition, another reference $ref2$ is created that points to the same object as $ref\_p$
}
\label{fig-ejemplo-2c}
\end{figure}
\end{example}

\subsubsection{Complex Types ($array$ type of objects with primitive attributes and with object attributes)}

We begin by explaining the visual representation of the array type, which contains objects with primitive attributes. This type of structure is declared in Java using the syntax \verb+Object[] array = new Object[N]+; where (\verb+N+) represents the number of elements that will make up the array. Each position in the array acts as an individual reference to an object, and these objects are composed exclusively of primitive type attributes.

From a visual perspective, a main reference—called $ref\_obj$—is modelled, pointing to a memory block containing an array of references. Each of these references, in turn, points to a different object. Since these objects are composed solely of primitive attributes, they are considered terminal nodes within the representation; that is, they do not contain other internal references. This feature limits the complexity of the visual structure, allowing for a flat and easily interpretable hierarchical representation.

\begin{figure}[ht]
\centering
\includegraphics[width=10.5cm]{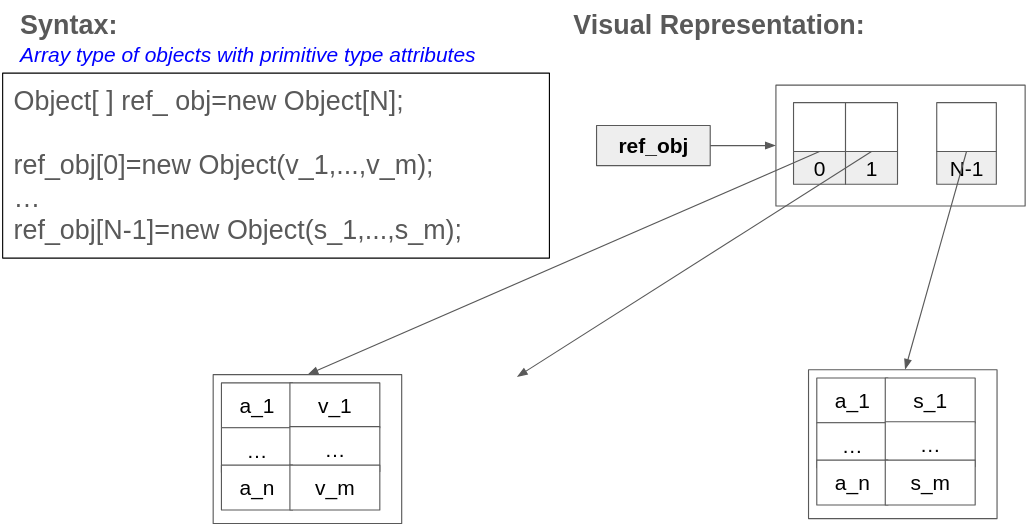}
\caption{Visual representation of the $array$ type of objects with primitive type attributes.}
\label{fig-representacion-complejos1}
\end{figure}

\begin{example}
Let's assume the $Person$ class is used and we declare an $array$ of size two Persons. Two references will then be created, pointing to their respective objects. In this case, we can see that the reference "$array\_personas[0]$" points to an object of type Person consisting of the string "000" and the age 56.

\begin{figure}[ht]
\centering
\includegraphics[width=10.5cm]{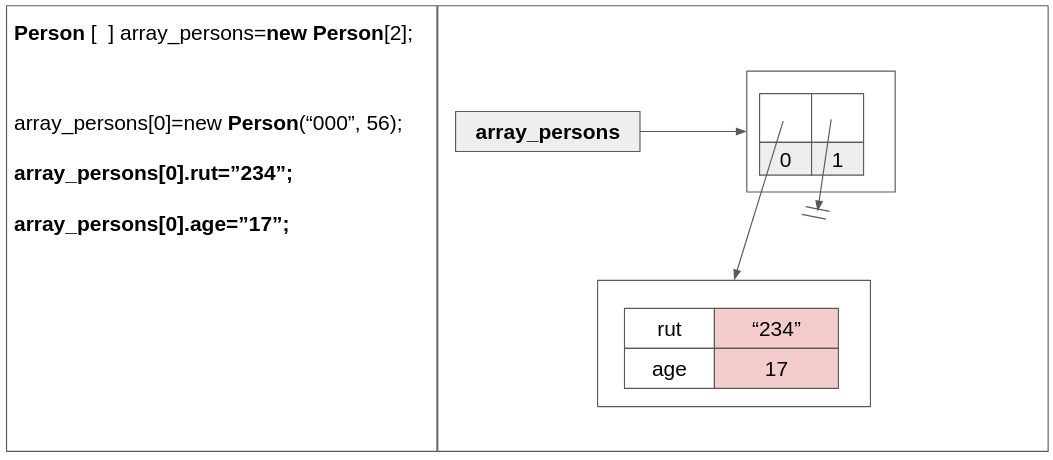}
\caption{Visual simulation of the execution of the code shown on the left side of the image.}
\label{fig-representacion-ejemplo1-complejos}
\end{figure}
 
\end{example}

\textbf{Object Type with Object-Type Attributes.} In a program, an object may exist that has references to other objects. We can denote a reference as "$ref\_obj$," which points to a memory area that may contain other attributes that are also references. Note that this can continue indefinitely until one of the objects has attributes of a primitive type. In other words, we can scale an object through references, creating more complex structures. In fact, this type is fundamental for the creation of data structures such as linked lists, trees, and graphs. For this reason, these representations can also be used in more advanced courses, such as data structures and artificial intelligence.

\begin{figure}[ht]
\centering
\includegraphics[width=10.5cm]{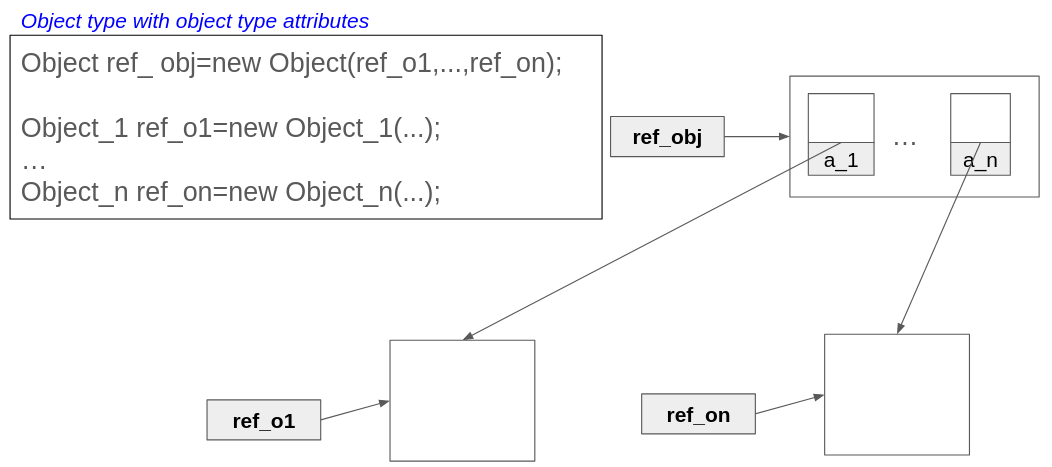}
\caption{Visual representation of the object type with object type attributes}
\label{fig-representacion-complejo2}
\end{figure}

\begin{example}
Let's suppose we have the class $Person$ (as seen previously) and the class $Friends$, which consists of two references of type $Person$ called $p1$ and $p2$. This relationship represents that two people are friends. When the main function is executed, $ref\_p1$ points to a memory area where the object $Person$ with ID “$234$” and age $56$ is located, while the reference $ref\_p2$ points to a memory area where the object with $rut$ having the value “$134$” and age $46$ is located. Additionally, we have the reference $a1$, which points to a memory area containing an object of type $Friends$ that references the same area as $ref\_p1$ and the same area as $ref\_p2$ (see Figure \ref{fig-representacion-ejemplo2-complejos})

\begin{figure}[ht]
\centering
\includegraphics[width=10.5cm]{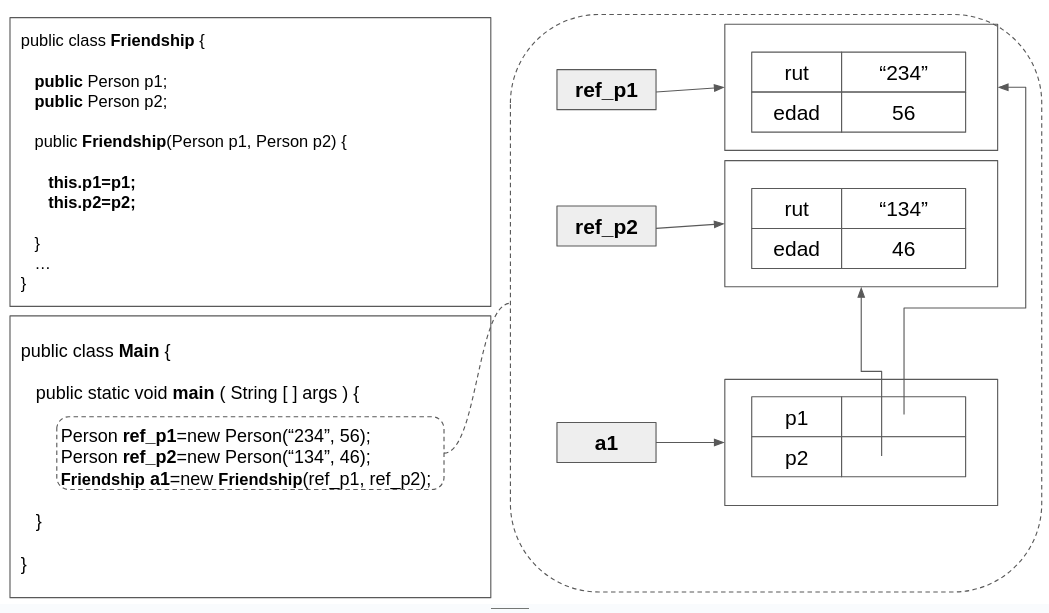}
\caption{Visual simulation of the execution of the code shown in the main method on the left side of the figure}
\label{fig-representacion-ejemplo2-complejos}
\end{figure}
\end{example}

\subsection{Preliminary survey of students}

To understand students' perceptions of the usefulness of visual representations and simulations of program execution in Java, we designed a brief survey consisting of three questions. This tool facilitates the collection of both qualitative and quantitative feedback on the effectiveness of this strategy in teaching key concepts of object-oriented programming.

\begin{enumerate}
\item Considering the visual representation of object creation, please answer the following: Was the code visualization helpful in understanding the concepts of objects and references in the Object-Oriented Programming (OOP) course?
\item What grade did you receive in the Object-Oriented Programming (OOP) course?
\item Please provide a general comment or opinion on the usefulness of the visualization during the learning process.
\end{enumerate}

\begin{figure}[ht]
\centering
\includegraphics[width=8.5cm]{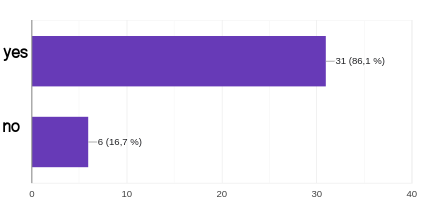}
\caption{Graph showing the percentage of male and female students who find visual representation useful.}
\label{fig-encuesta}
\end{figure}

The survey results are presented in Figure \ref{fig-encuesta} and Table \ref{tab:encuesta}. A total of 36 students responded to the questionnaire. Of these, 86.1\% indicated that code visualization was useful for understanding object and reference concepts in the Object-Oriented Programming course. In contrast, 16.7\% indicated that they did not find this tool useful for this purpose.

Many of the comments are very positive and demonstrate the usefulness of visual representations for understanding program execution. However, there are also some valuable, albeit less positive, comments that could help improve the proposal. Of these, we can highlight the following:

\begin{enumerate}
\item The need to standardize both representations and visual simulations of program execution to avoid ambiguities and facilitate understanding.
\item The importance of considering different learning styles: while visual representation is useful for many students, it is also important to incorporate verbal and explanatory elements.
\item The convenience of introducing this strategy at the beginning of the course, clearly explaining what it entails and how it should be interpreted, as some students expressed confusion about it.
\end{enumerate}

\begin{figure}[ht]
\centering
\includegraphics[width=8.5cm]{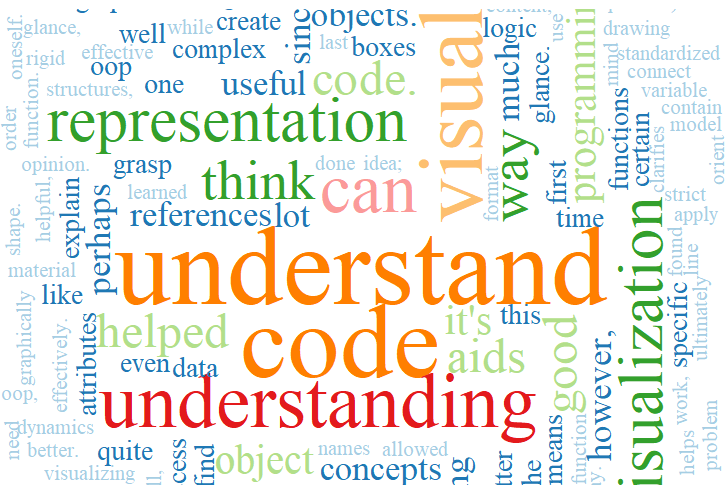}
\caption{Word cloud about student opinions.}
\label{fig-nube}
\end{figure}

\begin{figure}[ht]
\centering
\includegraphics[width=8.5cm]{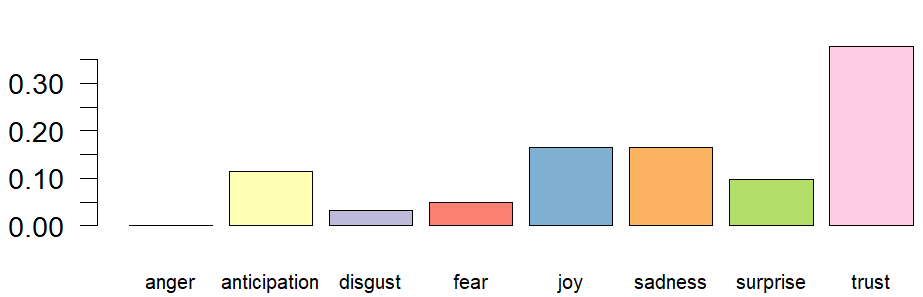}
\caption{Chart on emotions associated with students' opinions.}
\label{fig-emociones}
\end{figure}

The results of this analysis can be seen in Figures \ref{fig-nube} and \ref{fig-emociones}. Figure \ref{fig-nube} shows a word cloud generated from student comments, in which terms such as $visual$, $representation$, $helped$, $understand$, $understanding$ and $code$ stand out, suggesting a positive assessment of the tool. For its part, Figure \ref{fig-emociones} presents an analysis of the emotions detected in the responses, where the emotion of trust predominates, although some negative emotions are also evident, possibly associated with experiences of confusion or initial difficulty.

\begin{table*}[ht]
\begin{center}
\begin{tabular}{ |c|c|c|p{13.5cm}| }
\hline
 ID & YES/NO & GRADE & COMMENTS AND OPINIONS \\ \hline
1 & NO & 2.7 & You can understand the code represented in this visual format at a glance. \\\hline
2 & YES  &  Sin nota &  No comments \\  \hline
3 & YES & 4.4 & It is okey \\  \hline
4 & YES & 7.0 & I found it much easier to grasp the concepts using boxes and arrows, as they provided a clearer visual representation that allowed me to comprehend the code more effectively. \\  \hline
5 & YES & 4.4 & I believe this is a great idea; visual representation always aids in understanding the problems being implemented. \\  \hline
6 & YES & 7.0 & Visualizing code enhances comprehension for individuals who learn better through visual means rather than through reading. \\  \hline
7 & YES & 1.9 &  It aids in a deeper understanding of the code  \\  \hline
8 & YES & 4.2 &   \\  \hline
9 & NO & 4.1 & While introducing visual representation was helpful, I think it's even more valuable in subsequent courses, such as data structures, in my opinion.  \\  \hline
10 & NO & 4.3 &  Overall, it can be difficult to understand and explain complex concepts.  \\  \hline
11 & YES & 6.3 & In my opinion, it was very useful when implementing code because I could keep in mind the references I could access when using them. It clarifies the dynamics of what is presented.  \\  \hline
12 & YES & 4.2 &  It clarifies the dynamics of what is presented. With these examples, I learned what it means to create objects.  \\ \hline
13 & YES & sin nota & I find it an effective way to resolve and understand references while creating objects.   \\  \hline
14 & YES & 4.0 & Perhaps more explicit variable names would improve understanding. Still, the material is helpful as it is. \\  \hline
15 & YES & 4.9 &  It's a good way to understand object-oriented programming; however, it is a method that needs to be standardized so that everyone can understand a representation and create one in the same way.  \\  \hline
16 & YES & 6.9 & It is very practical for graphically explaining how references work, allowing us to imagine and understand the logic behind the code much better. This is ultimately the goal of the course: to grasp the logic of how object-oriented programming works so we can apply it to problem-solving. Personally, it helped me understand much more simply how a line of code functions when it calls a function and delivers attributes or how to access them. \\  \hline
17 & YES & 4.3 & Perhaps the model, at first glance, tends to explain the program simply; however, when you first look at the content, it can be confusing to connect the visual section with the code section. It could be simplified further by not being so strict with the diagram and drawing it in a more free-form, less rigid shape. (However, this model is quite simplified and precise.)  \\  \hline
18 & YES  & 6.8 & It's perfectly understandable; I even remembered a bit of the material. 
 \\  \hline
19 & YES & 4.5 &  I think it's good practice since visualization significantly aids in understanding concepts like references. I feel that visual representations of how certain structures work are very useful for grasping how they actually function.  \\  \hline
20 & YES & 6.1 & I believe that representation and visualization, whether for object programming or other programming languages, are essential and fundamental for understanding what is happening in the code. Furthermore, when solving problems, creating diagrams and visual representations helps generate greater mental order to find a solution to the problem and better understand the language and paradigm. \\  \hline
21 & YES  & 5.2 & No comments \\  \hline
22 & YES & 7.0  & Visualization helped me a lot in understanding the concept of objects. I think visualization is beneficial as a way of justifying code. Likewise, understanding what the code does doesn't necessarily have to be through a representation. \\  \hline 
23 & YES  & 6.9 &  This would be very helpful for people who cannot visualize the code at a glance. In addition to complementing the knowledge learned \\  \hline
24 & YES & 4.0 &  It is sufficiently informative and explanatory. However, a more visual example of creating a "person" object might be more appealing to students. 
\\  \hline
25 & YES & 4.3 & No comments \\  \hline
26 & YES & 4.5 &  The reference is well done; what is being demonstrated is quite clear, and it reminds me a lot of the visual aids at \verb+https://pythontutor.com/+. \\  \hline
27 & YES & 4.3 & It's a representation that helped me understand what each object should contain and how they should be connected.  \\  \hline
28 & YES & 3.5 & It's a good way to orient oneself.  \\  \hline
29 & YES & 4.3 & No comments \\  \hline
30 & YES & 5.3 & The use of boxes and arrows in the visualization greatly aids in understanding how classes and OOP code work in general, in a simple and easy-to-understand manner. Personally, the graphical explanations of code functions have helped me a lot to understand them, both in OOP and in Data Structures.  \\  \hline
31 & YES & 3.8 & I think it's a good way to learn OOP, and the teacher explains it very clearly.  \\  \hline
32 & YES & Without grade & No comments \\  \hline
33 & YES & 5.2 & I think that visualization can be useful to aid understanding in certain types of people, or in some exercises, examples or when explaining, as I answered previously for me it was a help especially at the beginning of the subject, then with the passage of time I started to leave it aside since I was able to understand what I was doing without the need for visualization. \\  \hline
34 & NO & 3 & If you show this to someone who has never studied programming in their life, they won't understand. \\
\hline
35 & YES & 5.6 & It helped me understand the creation of objects with their attributes and how they were joined together.  \\\hline
36 & YES & 6.5 & I think visualization is a good way to justify the code. Similarly, understanding what the code does does not necessarily have to be done through a graphical representation. It can also be more educational if you can understand the code in other ways, whether it be a specific explanation written in your own words, or something striking to the visualization beyond a diagram with rectangles and arrows. In my opinion, at the time I saw the visual representation in the last form named, I simply could not compare it well with the code. It was more difficult for me than a specific explanation of it since it made it complex to understand where one object referred to another and how they passed through inheritance or other methods.  \\\hline
\end{tabular}
 \caption{Responses to the survey regarding the usefulness of visual representations.}
  \label{tab:encuesta}
\end{center}
\end{table*}

\section{Conclusions and Future Works}
\label{sec-final}

This article addresses the revolution that code generators based on large language models represent in the field of programming and its teaching-learning processes. The main advantages and disadvantages of using these tools in introductory programming courses are listed, with a fundamental disadvantage highlighted: these systems lack true understanding of the code they generate, given that they are based on statistical models. Furthermore, the potential negative repercussions of excessive use of these technologies on student autonomy are discussed.

In response to this challenge, an expansion of both the content and assessment mechanisms has been proposed, emphasizing deep code comprehension. After reviewing the literature, the visual simulation of program execution was identified as a promising alternative, which was adapted for application in an introductory Java programming course. The main visual representations and the simulation dynamics associated with them were presented. Finally, a qualitative analysis of student opinions was conducted, highlighting terms such as "visual", "representation", "helped," "understand," "understanding," and "code" in the word cloud, along with the predominant emotion of trust detected in the comments.

This initial proposal paves the way for future research and development. Moving forward, we plan to create a tool that automates the generation of visual representations and incorporates an automatic feedback module in natural language. This module will allow for comparisons between the representations generated by students and those generated by the system, thus facilitating the monitoring and tracking of the level of understanding of the concepts taught in introductory programming courses. 

\section*{Declarations}

\subsection*{Ethics approval and consent to participate}

Not applicable.

\subsection*{Consent for publication}

Not applicable.

\subsection*{Availability of data and material}

Contact the corresponding author for data requests.

\subsection*{Competing interests}

Authors declare that they have no competing interests.

\subsection*{Funding}

This study was not supported by a grant.

\subsection*{Authors' contributions}

The study was designed by all the authors. Material preparation and data
collection were performed by all the authors. The data analysis was performed
by all the authors. The first draft of the manuscript was written by the first author, and all authors commented on previous versions of the manuscript. All authors read and approved the final manuscript.

\subsection*{Acknowledgments}

To all the students who anonymously answered the questionnaire 

\bibliographystyle{ieeetr}
\bibliography{bibliography}








\end{document}